\documentclass[aps,prb,floatfix,superscriptaddress,noshowpacs,twocolumn]{revtex4-2}

\usepackage{graphicx,amsmath,amssymb,epstopdf,wasysym}
\usepackage[space]{grffile}

\usepackage[hyperfootnotes=false]{hyperref}
\usepackage{bm}

\usepackage[dvipsnames]{xcolor}
\usepackage{braket,amsfonts}
\usepackage{dsfont}
\usepackage{float} 

\allowdisplaybreaks

\newcommand{\be}{\begin{equation}}
\newcommand{\ee}{\end{equation}}
\newcommand{\bga}{\begin{gather}}
\newcommand{\ega}{\end{gather}}
\newcommand{\bea}{\begin{eqnarray}}
\newcommand{\eea}{\end{eqnarray}}
\newcommand{\dagga}{{\phantom{\dagger}}}

\newcommand{\dis}{\displaystyle}

\newcommand{\up}{\uparrow}
\newcommand{\down}{\downarrow}
\newcommand{\fract}[2]{\frac{\dis \;#1\;}{\dis \;#2\;}}
\newcommand{\Tr}{\mathrm{Tr}}
\newcommand{\eqn}[1]{(\ref{#1})}

\newcommand{\bw}{\begin{widetext}}
\newcommand{\ew}{\end{widetext}}
\newenvironment{eqs}%
{\begin{equation} \begin{aligned}}%
{\end{aligned} \end{equation} }
\newcommand{\beal}{\begin{eqs}}
\newcommand{\eal}{\end{eqs}}

\newcommand{\bealn}{\beal\nonumber}
\newcommand{\coloneq}{\mathrel{\mathop:}\mathrel{\mkern-1.2mu}=}
\footnotetext{These authors contributed equally to this work.}

\begin{document}
\title{Direct minimization versus iterative embedding in the ghost-Gutzwiller method: \\ a comparative study of magnetism in Mott insulators}
\author{Antonio Maria Tagliente$^\S$}
\email{atagliente@sissa.it}
\affiliation{International School for Advanced Studies (SISSA), Via Bonomea 265, I-34136 Trieste, Italy}

\author{Ivan Pasqua$^\S$}
\email{ipasqua@sissa.it}
\affiliation{International School for Advanced Studies (SISSA), Via Bonomea 265, I-34136 Trieste, Italy}

\author{Michele Fabrizio}
\affiliation{International School for Advanced Studies (SISSA), Via Bonomea 265, I-34136 Trieste, Italy} 

\begin{abstract}
Accurately describing a hypothetical symmetry-invariant Mott insulator presents a longstanding challenge in iterative quantum embedding methods. We address this issue within the ghost-Gutzwiller method, which can be solved either through an iterative embedding scheme, analogous to dynamical mean-field theory, or by directly minimizing its variational energy functional. Across the Mott transition of the single-band Hubbard model, these formally equivalent approaches behave very differently: the iterative scheme is computationally efficient but fragile, necessitating ad-hoc recipes in the Mott phase that fail in a Zeeman field, leading to a discontinuous energy and a spurious fully-polarized insulator. Direct minimization avoids these artifacts, stabilizing a genuinely paramagnetic solution. Conversely, when symmetry breaking is allowed, as in an antiferromagnetic phase, the iterative scheme yields the correct solution, closely aligning with dynamical mean-field theory. Our findings delineate the conditions under which the iterative embedding can be trusted and when direct minimization is instead required.

\end{abstract}

\maketitle
\vspace{-1em}

\section{Introduction}
The current theoretical understanding of correlated phenomena in condensed matter physics 
is grounded in self-consistent methods where, typically, the many-body problem is approximated by a simpler, non-linear model that is often solved iteratively, such as the Hartree-Fock approximation. A broad class of self-consistent methods known as quantum embedding theories (QE) \cite{QE} partition the original complex problem into a fragment, referred to as the \textit{impurity}, and its environment that are treated at different levels of accuracy, with the impurity being analyzed with greater precision. The coupling between environment and impurity is enforced by particular self-consistency conditions that distinguish the various embedding theories.\\
A paradigmatic quantum embedding method is Dynamical Mean Field Theory (DMFT) \cite{DMFT}. DMFT is exact in lattices with infinite coordination number and with local interactions, but it serves as a sensible approximation in realistic lattices, providing insightful results about correlated phenomena, particularly the Mott transition. Although DMFT can be formulated through an appropriate functional whose saddle point yields the single-particle Green's function \cite{Kotliar1999,ggutdmft}, in practice, the self-consistency equation is commonly solved  iteratively. The iteration scheme is computationally very convenient, but hinders the enforcement of symmetries that are spontaneously broken in the actual low-temperature state. 
A prominent example is the paramagnetic Mott insulating state of the half-filled Hubbard model, which stands as one of the key achievements of DMFT. Indeed, to prevent antiferromagnetism, which is known to occur at low temperatures, one must force the convergence of the iterative method through ad hoc  recipes of spin symmetrization. 
This is equivalent to employing a mixed state instead of a pure one, and may be responsible 
for the residual $\ln 2$ entropy per site of the $T=0$ Mott insulator found in DMFT.
Conversely, one might anticipate that the entropy of the free local moments could be quenched at zero temperature, even in the absence of symmetry breaking, due to the formation of a 
hypothetical spin-liquid state.\\
\noindent
Conventionally, spin liquids are described using Heisenberg-type Hamiltonians, where the charge 
is frozen right from the outset. However, the substantial number of candidate spin-liquid materials \cite{spin-liquid-review-1,spin-liquid-review-2,Matsuda_2025} makes it worthwhile to explore the interplay between charge and spin dynamics. Quantum embedding theories, such as DMFT, hold genuine potential to address this issue. Nevertheless, 
these methods currently struggle to account for the strong frustration and massive spin entanglement \cite{Savary_2017} that characterize spin-liquids, which can only be described by artificially forcing $SU(2)$ symmetry. 
As mentioned earlier, this enforcement is challenging in iterative schemes.  
Here, we show how this problem can be circumvented in an alternative quantum embedding method, the so-called ghost-Gutzwiller approximation (ghost-GA) \cite{Nicola-ghost}. On infinite-coordination lattices, ghost-GA provides a strictly variational estimate of the ground-state energy, otherwise acting as a
legitimate approximation in realistic lattices just like DMFT. This connection has recently been strengthened by the demonstration that ghost-GA becomes equivalent to DMFT in the limit of infinitely many ghost orbitals \cite{ggutdmft}. 
Remarkably, already with few auxiliary orbitals, ghost-GA offers a computationally efficient and complementary description of strongly correlated systems \cite{Nicola,Guerci2019,frank2021,Carlos2023,gRISB,gRISB-DFT}, whose versatility has enabled applications to a broad range of phenomena~\cite{Carlos2024, tagliente2025spinons, arXivtrilayer2025, giuli2025altermagnetism, pasqua2026topology, gabriele, Carlos2026, giuli-foundation, Chen2026, kazemi2026magnetism, pasqua2026FLstar}.

\section{Direct minimization vs iterative embedding}
\noindent
The ghost-Gutzwiller wavefunction and the corresponding approximation have been extensively discussed in the literature, both in the original work \cite{Nicola-ghost} and in a recent account \cite{ggutdmft}. The purpose of this section is therefore not to rederive the ghost-GA, but rather to highlight the key steps that connect the variational formulation to its embedding representation, which closely resembles DMFT. In the following sections, we will argue that this embedding scheme, together with its iterative solution, suffers from the same difficulties previously identified in DMFT. By contrast, we will show that a direct variational calculation avoids these issues.\\
\noindent
The ghost-GA is built on a variational wavefunction ansatz
\beal
\ket{\Psi} = \mathcal{P} \ket{\psi_*} = \prod_i \mathcal{P}_i \ket{\psi_*}\,,
\label{GW-wavefunction}
\eal
where the uncorrelated $\ket{\psi_*}$ is a Slater determinant defined in an enlarged auxiliary Hilbert space of $N_\text{aux}$ orbitals, as opposed to the $N_\mathrm{phys}$ physical ones; here and throughout, the spin degree of freedom is included in the orbital count. The difference $N_\mathrm{ghosts}\equiv N_\text{aux}-N_\mathrm{phys}$ is conventionally referred to as the number of ghosts. The variational linear operator $\mathcal{P}_i$ in \eqn{GW-wavefunction} maps local configurations of the auxiliary fermions into those of the physical ones, namely, 
\beal
\mathcal{P}_i  = \sum_{\Gamma_\text{phys} \, \gamma_\text{aux}} \Lambda(i)_{ \Gamma_\text{phys}\gamma_\text{aux}} \ket{\Gamma_\text{phys}, i} \bra{\gamma_\text{aux}, i}\, , \label{linearmap}
\eal
with variational parameters $\Lambda(i)_{\Gamma_\text{phys}\gamma_\text{aux}}$, in general site-dependent, and where $\ket{\Gamma_\text{phys}, i}$ and $\ket{\gamma_\text{aux},i}$ denote the bases of the physical and auxiliary Hilbert spaces, respectively, at site $i$. We henceforth denote the physical fermionic operators at site $i$ through the $N_\mathrm{phys}$-component spinors $\mathbf{c}_i$ and the auxiliary ones through the $N_\text{aux}$-component spinors $\mathbf{f}_i$. We consider a generic Hamiltonian 
\begin{equation}
    H = \sum_{i\not=j}\, \mathbf{c}^\dagger_{i}\, \hat{t}_{ij} \, \mathbf{c}^\dagga_{j} + 
\sum_i\, H_\mathrm{loc}\big[\mathbf{c}^\dagger_{i},\mathbf{c}^\dagga_{i}\big]\,,
\label{multibands}
\end{equation}
where the matrix $\hat{t}_{ij}$ encodes the hopping amplitudes between different sites $i$ and $j$, while $H_\mathrm{loc}$ collects the one- and two-body local terms. In infinite-coordination lattices, the expectation value of $H$ over the variational wavefunction \eqn{GW-wavefunction} can be evaluated analytically, provided the constraints
\begin{equation}
    \begin{aligned}
    \bra{\psi_*} \mathcal{P}_i^\dagger\,\mathcal{P}_i^\dagga \ket{\psi_*}
    &= 1\,,\\
    \bra{\psi_*} \mathcal{P}_i^\dagger\,\mathcal{P}_i^\dagga\,
    \mathbf{f}_i^{\dagga}\otimes \mathbf{f}_i^\dagger \ket{\psi_*}
    &= \bra{\psi_*} 
    \mathbf{f}_i^{\dagga}\otimes \mathbf{f}_i^\dagger \ket{\psi_*}\\
    &\coloneq \hat{\Delta}_{*}(\psi_*, i)\,,
    \label{GA-constraint}
    \end{aligned}
\end{equation}
are satisfied. Applying the same formulas when the coordination number is finite defines the Gutzwiller approximation, hence our notation ghost-GA. In what follows we suppress the site index and present the formalism for a translationally invariant state, where all local quantities are site-independent; the generalization to a broken-symmetry state, where these quantities acquire a sublattice dependence, is straightforward and is used explicitly in Sec. IV.\\
Following~\cite{Nicola}, the parameters $\Lambda_{\gamma_\text{phys}\gamma_\text{aux}}$ can be identified with the components of an impurity wavefunction 
\begin{equation}
\ket{\phi}  = \sum_{\Gamma_\text{phys}\bar{\gamma}_\text{aux}} \phi_{ \Gamma_\text{phys}\bar{\gamma}_\text{aux}} \ket{\Gamma_\text{phys}} \ket{\bar{\gamma}_\text{aux}}\,,
\label{impurity-WF}
\end{equation}
with $\bar{\gamma}_\text{aux}$ related to $\gamma_\text{aux}$ by an appropriate particle-hole transformation \cite{Nicola}. The wavefunction \eqn{impurity-WF} describes an impurity carrying the same degrees of freedom as the physical fermions, coupled to $N_\text{aux}$ baths; accordingly, we denote the impurity annihilation operator by the spinor $\mathbf{c}$ and the bath one by $\mathbf{f}$, now without a site index. Given any physical local multi-body operator $\mathcal{O}$, 
one can demonstrate within the ghost-GA the following equivalence 
\begin{equation}
\bra{\Psi} \mathcal{O}\big[\mathbf{c}^\dagger_{i},\mathbf{c}^\dagga_{i}\big] \ket{\Psi} \equiv \bra{\phi} \mathcal{O}\big[\mathbf{c}^\dagger,\mathbf{c}^\dagga\big] \ket{\phi}\, .
\label{local}
\end{equation}
Moreover, the constraints \eqn{GA-constraint} become 
\begin{equation}
    \begin{aligned}
    \braket{\phi|\phi} &= 1\,,\\
    \hat{\Delta}_*(\psi_*) &= 
    \mathbb{I} - \bra{\phi} 
    \mathbf{f}\otimes \mathbf{f}^\dagger \ket{\phi} 
    \coloneq \mathbb{I} -\hat{\Delta}_\phi(\phi)
    \,,
    \end{aligned}
    \label{GA-constraint-bis}
\end{equation}
with $\mathbb{I}$ the identity matrix. The first equation in \eqn{GA-constraint-bis} simply imposes the normalization of the impurity wavefunction; the second requires the local one-body density matrix of the Slater determinant $\ket{\psi_*}$ to match the particle-hole transform of the bath one-body density matrix of the impurity wavefunction. Since \eqn{GW-wavefunction} exhibits gauge symmetry, $\mathcal{P}_i\to \mathcal{P}_i\,U_i^\dagger$ and $\ket{\psi_*}\to \prod_i U_i \ket{\psi_*}$, with unitary $U_i$ whose generators are one-body operators that leave $\ket{\psi_*}$ a Fock state, one may take both density matrices to be diagonal without any loss of variational freedom, what is known as the natural basis.\\  
The final ingredient necessary to compute the variational energy functional is the expectation value over \eqn{GW-wavefunction} of the hopping term in \eqn{multibands}. Within ghost-GA one finds that 
\begin{equation}
\bra{\Psi}\mathbf{c}^\dagger_{i}\, \hat{t}_{ij} \, \mathbf{c}^\dagga_{j}\ket{\Psi}
= \bra{\psi_*}\mathbf{f}^\dagger_{i}\, \hat{R}(\phi)^\dagger\,\hat{t}_{ij} \,\hat{R}(\phi)\, \mathbf{f}^\dagga_{j}\ket{\psi_*}\,,
\label{hopping-expectation}
\end{equation}
where \cite{michele-fluctuations} 
\begin{equation}
    \hat{R}(\phi) =    \hat{Q}(\phi)\cdot \hat{S}^{-1}(\phi) \,,\label{R}
\end{equation}
with
\begin{equation}
    \begin{aligned}
        \hat{S}(\phi)^2 &= \hat{\Delta}_\phi(\phi) \cdot \big( \mathbb{I} - \hat{\Delta}_\phi(\phi)\big)\,,\\
        \hat{Q}(\phi) &= \bra{\phi} \mathbf{c}\otimes \mathbf{f}^{\dagger}\ket{\phi}\,. \label{QS}
    \end{aligned}
\end{equation}
Collecting these results, the energy per site of \eqn{multibands} over the variational wavefunction \eqn{GW-wavefunction}, subject to the constraints \eqn{GA-constraint-bis}, reads
\beal
\mathcal{E}(\psi_*, \phi) &=  \fract{1}{V}\sum_{i\not=j}\,\bra{\psi_*}\mathbf{f}^\dagger_{i}\, \hat{R}(\phi)^\dagger\,\hat{t}_{ij} \,\hat{R}(\phi)\, \mathbf{f}^\dagga_{j}\ket{\psi_*}\\
&\qquad  + \bra{\phi} H_{loc}\big[\mathbf{c}^\dagger,\mathbf{c}^\dagga\big] \ket{\phi}\,, \label{energy functional}
\eal
with $V$ the number of sites. 
Equivalently, enforcing the constraints \eqn{GA-constraint-bis} through 
a Lagrange multiplier matrix $\hat{\lambda}$, and the normalizations of $\ket{\psi_*}$ 
and $\ket{\phi}$ by two additional Lagrange multipliers $E_*$ and $E_\phi$, respectively, 
the minimization of \eqn{energy functional} 
translates to finding the saddle point of the functional
\begin{equation}
    \begin{aligned}
    \mathcal{F} &=  \mathcal{E}(\psi_*, \phi) - E_* \braket{\psi_*|\psi_*} - E_{\phi} \braket{\phi|\phi}  \\
    &\qquad  + \Tr{ \Bigg[\hat{\lambda}\cdot
        \Big\{ \mathbb{I} -\hat{\Delta}_\phi(\phi) - \hat{\Delta}_*(\psi_*)
        \Big\} \Bigg]}\,.
    \end{aligned}
    \label{functional}
\end{equation}
The saddle-point equation for $\ket{\psi_*}$ is just the ground-state problem of a non-interacting Hamiltonian whose parameters depend on $\ket{\phi}$ and $\hat{\lambda}$. The equation for $\ket{\phi}$, by contrast, is a non-linear Schrödinger equation, since the functional derivative of $\hat{R}(\phi)$ in \eqn{R} is not linear in $\ket{\phi}$. Even so, one can always adopt a suitable parametrization of $\ket{\phi}$ and search directly for the saddle point of \eqn{functional} without ever solving the non-linear Schrödinger equation; we will refer to this as the \textit{direct minimization}.\\
An alternative route to the stationary point of the functional \eqref{functional} is the \textit{iterative embedding scheme}, in which additional Lagrange multipliers turn the equation for the impurity wavefunction into a linear eigenvalue problem, so that $\ket{\phi}$ becomes the ground state of an impurity model. 
Specifically, one introduces the Lagrange multiplier matrix 
$\hat{V}$ to enforce equation \eqn{R}, while treating $\hat{R}$ as an independent parameter. Furthermore, a variational density matrix $\hat{\Delta}$ is introduced, together with the Lagrange multiplier matrices $\hat{\lambda}$ and $\hat{\lambda}^{c}$ that force $\hat{\Delta}_*(\psi_*)$ and $\hat{\Delta}_\phi(\phi)$ to be equal to $\hat{\Delta}$. The functional to be extremized then becomes
\beal
    \mathcal{L} &=  \mathcal{E}'(\psi_*, \hat{R}) - E_* \braket{\psi_*|\psi_*} - E_{imp} \braket{\phi|\phi}  \\
    &\qquad + \Tr{\Bigg[  \hat{\lambda}\cdot \Big(\hat{\Delta}_*(\psi_*)-\hat{\Delta}\Big)\Bigg]}\\
    &\qquad 
    - \Tr{\Bigg[\hat{\lambda}^{c}\cdot\Big(\mathbb{I} - \hat{\Delta}_\phi(\phi) -\hat{\Delta}\Big)\Bigg]}\\
    &\qquad + \Tr{\Bigg[ \bigg\{\Big( \hat{R}\cdot  \hat{S}(\hat{\Delta})  -  \hat{Q}(\phi)\Big) \cdot \hat{V} + H.c.\bigg\}\Bigg]}
    \,,
    \label{Lagrangian}
\eal
where $\mathcal{E}'(\psi_*, \hat{R})$ is derived from $\mathcal{E}(\psi_*, \phi)$ in 
\eqn{energy functional} replacing $\hat{R}(\phi)$ with $\hat{R}$, while $\hat{S}(\hat{\Delta})$ is obtained from the first line in equation \ref{QS} replacing $\Delta_{\phi}( \phi )$ with $\Delta$. 
After these manipulations, the saddle-point equations of \eqn{Lagrangian} can be cast as a set of self-consistency conditions that bear a very close resemblance to the DMFT self-consistency loop \cite{ggutdmft}, the significant difference being that they require only the ground state of the impurity model rather than its frequency-dependent Green's function. As in DMFT, this saddle point is then readily reached through iterative methods. \\
It should be stressed, however, that promoting the one-body density matrices $\hat{\Delta}$ and $\hat{R}$ in \eqn{Lagrangian} to free variational parameters is a non-trivial assumption, since these objects obey strong representability constraints \cite{Nrep}. As a consequence, the minimum of the variational energy \eqn{energy functional} may lie on the boundary of the region where the one-body density matrices are defined, rather than at a saddle point of \eqn{Lagrangian}. As we shall see, this is precisely what happens in the Mott insulating phase, where some entries of $\hat{R}$ vanish and the minimum is pushed to the boundary of the admissible region — a regime in which the embedding scheme becomes especially delicate. For this reason, in the following we compare, across several examples, the solution of the saddle-point equations of \eqn{Lagrangian} with the result of a direct optimization of the impurity-wavefunction coefficients in \eqn{impurity-WF} for the original variational functional \eqn{functional}.

\section{The paramagnetic Mott transition within ghost-Gutzwiller}
\begin{figure}[t]
    \centering
    \includegraphics[width=0.5\textwidth]{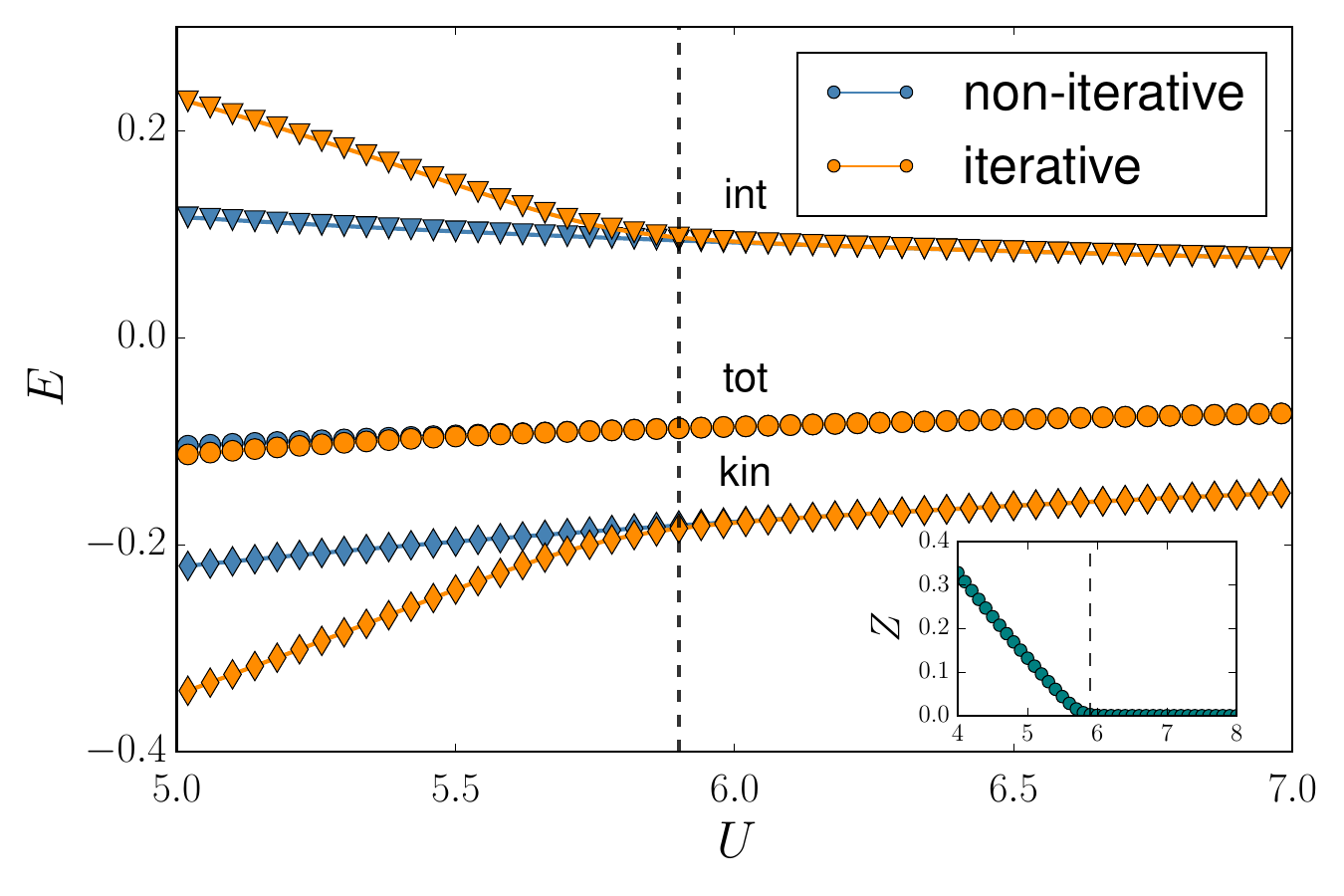}
    \caption{Comparison between the energies per site obtained with the iterative embedding scheme (\emph{iterative}) and with the direct minimization of the impurity wavefunction \eqn{phi Lanczos} at $\phi=\pi/4$ (\emph{non-iterative}): from top to bottom, local interaction energy (\textit{int}), total energy (\textit{tot}), and kinetic energy (\textit{kin}), as a function of $U$. For each of them, the two methods coincide above $U_c$ (dashed line), where the system is in the paramagnetic Mott state. Inset: quasiparticle residue $Z$ vs $U$, locating $U_c$.}
    \label{energiesPM}
\end{figure}

\noindent
We begin by examining the paramagnetic, translationally invariant solution of the single-band Hubbard model at half-filling and on a bipartite lattice. While the true ground state in this case is an antiferromagnetic insulator, it is still interesting to inquire about the potential description of a hypothetical metastable paramagnetic insulator, devoid of the complexities introduced by the frustration required to stabilize a genuine spin liquid. \\
The model Hamiltonian reads
\begin{align}
\mathcal{H} = -t \sum_{\langle ij \rangle \sigma}\, \big( c^{\dagger}_{i\sigma}\, c^\dagga_{j\sigma} + \mathrm{H.c.}\big) + \frac{U}{2}\sum_i\, (n_i - 1)^2~,
\end{align}
where $t=1$, our unit of energy, is the nearest-neighbor hopping and $U$ the on-site Coulomb repulsion. We consider a Bethe lattice with infinite coordination number, where the GA becomes exact. The non-interacting density of states is semicircular,
\begin{align}
    \mathcal{D}(\epsilon) = \frac{2}{\pi D} \; \sqrt{1 - \left(\frac{\epsilon}{D}\right)^2\;} \;,
\end{align}
with $D = 2$ in our units. Hereafter, we fix $N_\mathrm{ghosts}=2$, so that the impurity model describes an impurity site coupled to three bath sites.\\
We first consider the weakly correlated regime, $U<U_c$, where the system is a renormalized Fermi liquid. Restricting ourselves to the translationally invariant solution, the embedding scheme converges smoothly to a spin-unpolarized state, and the paramagnetic metal is a local minimum of the ghost-GA functional. Within this regime the impurity wavefunction is a non-degenerate singlet. This is the situation in which the embedding formulation works at its best, and replacing the explicit minimization of the functional by a self-consistent loop is both safe and computationally advantageous.\\
Upon increasing $U$, the system undergoes a Mott transition at a critical $U_c$ where the quasiparticle residue,
\begin{equation}
    Z = \Big( 1 - \partial_\omega \mathrm{Re} \Sigma(\omega)_{ | \omega=0} \Big)^{-1} \, ,
\end{equation}
where $\Sigma(\omega)$ is the self-energy on the real frequency axis, vanishes and the quasiparticle peaked at the chemical potential disappears in the spectral function. Correspondingly, in the impurity model the bath site sitting at the Fermi level, which for $U<U_c$ hybridizes with the impurity site, progressively decouples as $U\to U_c^-$ and is completely severed from the impurity for $U>U_c$. The impurity is then left coupled only to the two bath sites describing the lower and upper Hubbard bands, LHB and UHB, respectively. Technically, this decoupling shows up in the natural basis as the vanishing of one entry of the renormalization matrix $\hat R$; this feature is a hallmark of the Mott phase.
\begin{figure}[t]
    \centering
    \includegraphics[width=0.5\textwidth]{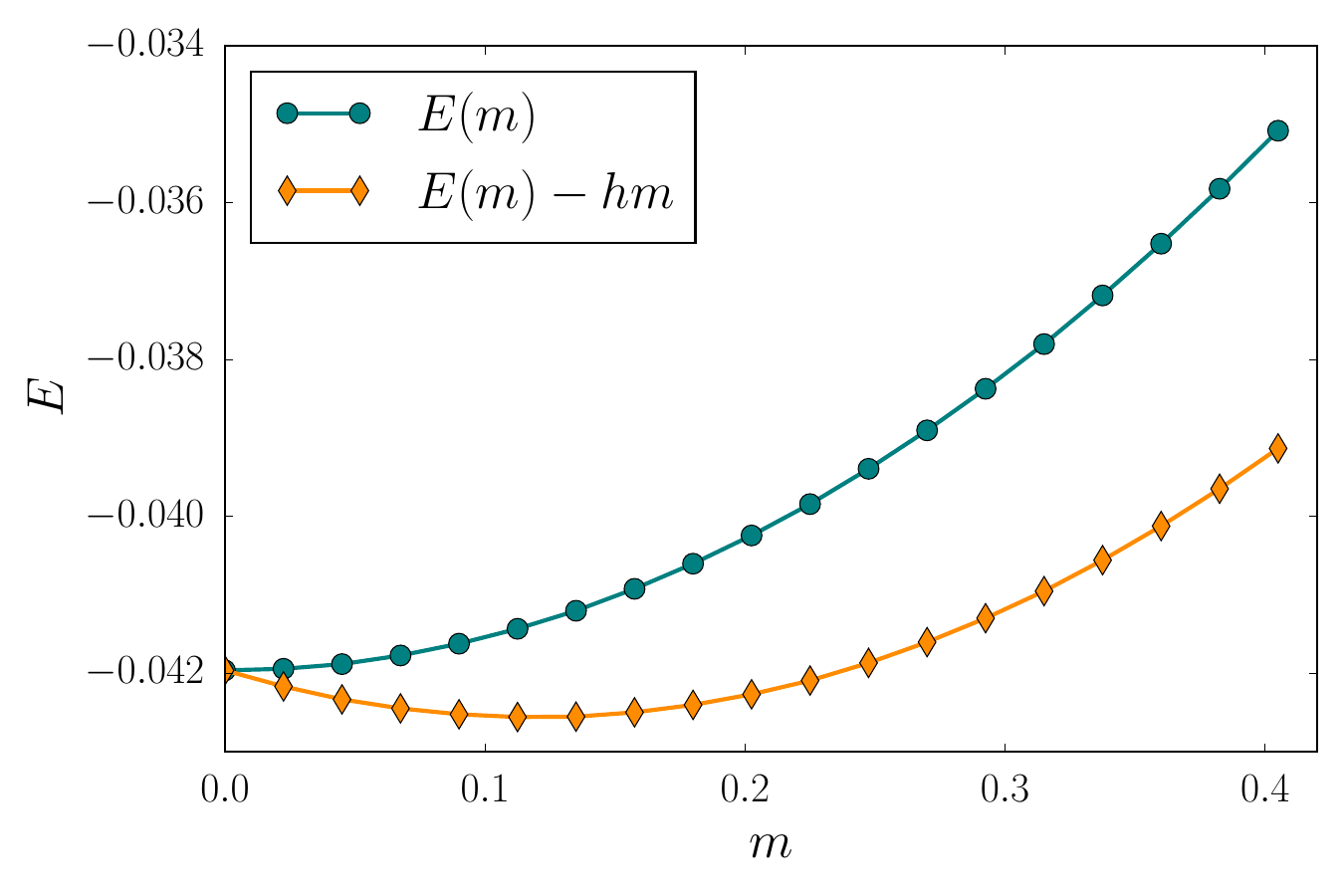}
    \caption{Ghost-GA energy $E(m)$ as a function of the fixed impurity magnetization $m$, for $U=12$ in the Mott phase, together with its Legendre transform $E(m)-hm$ in presence of a magnetic field $h=0.01$. $E(m)$ is minimum at $m=0$, confirming that the paramagnetic solution is a local minimum; switching on $h$ shifts the minimum to a finite magnetization.}
    \label{fig:Energy_vs_magnetization}
\end{figure}\\
\noindent
Once the bath at the Fermi level decouples, the impurity ground state becomes degenerate. This same degeneracy extends to the entire spectrum. Its origin lies in the gauge symmetry of the wavefunction \eqn{GW-wavefunction}: completely broken in the metal, it is restored in the Mott phase for the 
auxiliary fermion corresponding to the zero-energy decoupled bath. Specifically, if 
$\ket{\phi_\sigma}$, $\sigma=\up,\down$, is the spin-1/2 ground-state wavefunction of the impurity plus the two coupled baths, while   
$\ket{\gamma_1}$ and $\ket{\gamma_2}$ any two orthogonal states of the decoupled bath, then 
\bealn
\ket{\phi} &=  \ket{\phi_\up}\,\ket{\gamma_1} + \ket{\phi_\down}\,\ket{\gamma_2}\,,
\eal
yields the same variational energy for any choice of $\ket{\gamma_1}$ and $\ket{\gamma_2}$. 
Even if we partially fix the gauge by employing physical symmetries, such as  
imposing that $\ket{\phi}$ is eigenstate 
of the number of particles with eigenvalue corresponding to half-filling \cite{Nicola}, and of $S_z$ with eigenvalue $S_z=0$, thereby restricting the available states 
of the decoupled bath to $\ket{\gamma}=\ket{\up},\ket{\down}$, we are still left with a double 
degenerate impurity wavefunction   
\beal
\ket{\phi_\pm} &=  \ket{\phi_\up}\,\ket{\down} \mp \ket{\phi_\down}\,\ket{\up}\,,
\label{phi + and -}
\eal
where $\ket{\phi_+}$ is a spin singlet and $\ket{\phi_-}$ the $S_z=0$ component of the 
spin-triplet. The residual double degeneracy, which corresponds to the $Z_2$ gauge transformation $\ket{\up}\to -\ket{\up}$ that is recovered in the Mott phase and reminds what was discussed in \cite{Rok-PRB2015}, is not removable and is the root cause of a well-known difficulty encountered in the iterative embedding scheme within the Mott phase. This difficulty arises because the impurity ground state selected at each iteration is an uncontrolled linear combination of the two states in \eqn{phi + and -}, thereby hindering the convergence of the scheme, with the impurity magnetization flipping up and down from one iteration to the next. \\
A solution to this problem lies in explicitly resolving the degeneracy. Within the twofold degenerate manifold, any state of definite impurity magnetization $m$ can be constructed as a unique linear combination of the two states in \eqn{phi + and -} \cite{pasqua2026topology}. As $m$ approaches zero, the embedding loop converges smoothly to a spin-unpolarized solution. To verify that this is the physically correct choice, we show the ghost-GA energy $E(m)$ as a function of $m$ in Fig.~\ref{fig:Energy_vs_magnetization} for $U=12$ in the Mott phase. The energy is minimum at $m=0$, and it increase monotonically with $m$.\\
One might conclude that the Mott phase ambiguity is then solved. The prescription depicted above, however, works only because the two states are exactly degenerate: any perturbation that lifts the degeneracy spoils this construction and results in a completely polarized impurity ground state, at least within the embedding scheme we here consider. This is precisely what happens in a magnetic field, which we discuss in the next section, and it is expected to occur more generally whenever a perturbation splits the degenerate manifold, as for the crystal-field splitting or the Hund's coupling in multiband models. This motivated us to look for a more robust description of the Mott insulating phase within ghost-GA, i.e., directly minimizing the ghost-GA functional through a suitable parametrization of the impurity wavefunction $\ket{\phi}$ that encodes the projector $\mathcal{P}_i$.\\
To this end, we construct the generic impurity wavefunction with a singly-occupied decoupled bath, in the half-filled sector with $S_z=0$, and with the additional constraints of the coupled baths forming a natural basis, i.e., with a diagonal reduced density matrix, and spin unpolarized \cite{tagliente2025spinons}. 
This implies that only the impurity and the decoupled bath may carry magnetic moments, actually opposite to each other because of $S_z=0$. Practically, we start from the wavefunction 
\be
\ket{\phi_0} =  \cos\varphi\,\ket{\up,2,0}\ket{\down}
- \sin\varphi\,\ket{\down,2,0}\ket{\up}\,,
\label{phi-0} 
\ee
where $\ket{\gamma_\mathrm{imp},\gamma_\mathrm{LHB},\gamma_\mathrm{UHB}}\ket{\gamma_\mathrm{dec}}$, $\gamma=0,\up,\down,2$, are basis states, with $\ket{\gamma_\mathrm{dec}}=\ket{\up},\ket{\down}$ 
describing the decoupled bath. The angle $\varphi\in [0,\pi/2]$ parametrizes the impurity magnetization, which vanishes at $\varphi=\pi/4$. Next, we generate from \eqn{phi-0} the Lanczos chain by applying successively the hybridization between the impurity and the high-energy baths, LHB and UHB, forcing the above constraints. At the end, we obtain the impurity wavefunction 
\be
\ket{\phi} = \sum_{n=0}^4\, \alpha_n\ket{\phi_n}\,,
\label{phi Lanczos}
\ee 
with real $\alpha_n$ satisfying $\sum_n\,\alpha_n^2=1$, where 
\beal
\ket{\phi_1} &= \fract{1}{2}\,\Big(\ket{2,\down,0}\ket{\up}
-\ket{2,\up,0}\ket{\down}\\
&\qquad\qquad 
-\ket{2,2,\down}\ket{\up} + \ket{2,2,\up}\ket{\down}\Big)\,,\\
\ket{\phi_2} &= \fract{\sin\varphi}{\sqrt{6}}\,\Big(\ket{\up,\down,\up}\ket{\down}\,
+\ket{\up,\up,\down}\ket{\down}
\Big) \\
&\qquad + \fract{\cos\varphi}{\sqrt{6}}\,\Big(\ket{\down,\down,\up}\ket{\up} 
+\ket{\down,\up,\down}\ket{\up}
\Big)\\
&\qquad -\fract{1}{\sqrt{3}}\,\Big(\ket{\up,\down,\down}\ket{\up} 
+ \ket{\down,\up,\up}\ket{\down}\Big)\,,\\
\ket{\phi_3} &=\fract{1}{2}\,\Big(\ket{2,0,\up}\ket{\down} - \ket{2,0,\down}\ket{\up}\\
&\qquad\qquad - \ket{0,\up,2}\ket{\down} + \ket{0,\down,2}\ket{\up}\Big)\, \\
\ket{\phi_4} &= -\cos\varphi\ket{\up,0,2}\ket{\down} + \sin\varphi\ket{\down,0,2}\ket{\up}\, .
\label{phi 1-4}
\eal
Considering the normalization, the wavefunction \eqn{phi Lanczos} comprises five free parameters that are optimized by locating the saddle point of \eqn{functional}. Since \eqn{phi Lanczos} incorporates a decoupled bath, it is exclusively suitable for describing the Mott phase and not the correlated metal below $U_c$. \\
In Fig.~\ref{energiesPM} we compare the energies of the embedding scheme obtained through the limit 
$m\to 0$, see the above discussion, and the minimization of \eqn{phi Lanczos} directly at $m=0$, i.e., 
at $\phi=\pi/4$. As anticipated, the energies deviate below $U_c$, where \eqn{phi Lanczos} is not accurate, but coincide in the Mott regime, $U>U_c$. This both validates our compact ansatz \eqn{phi Lanczos} and confirms that the embedding solution, once the degeneracy is correctly resolved, captures the same physics.
This agreement holds only in the unperturbed paramagnet; in the next section, we show that a magnetic field, by lifting the degeneracy, drives the two schemes apart.

\subsection{The Mott transition in a magnetic field}
\noindent
A magnetic field that couples to the spin is the simplest perturbation that lifts the previously mentioned degeneracy in the Mott phase. It therefore provides a stringent test of the embedding scheme. We add to the Hamiltonian a Zeeman term,
\begin{equation}
H \to H - h \sum_i \big( n_{i\up} - n_{i\down}\big)\,,
\label{zeeman}
\end{equation}
and work at fixed magnetic field $h = 0.01$. An analogous problem was studied at fixed magnetization in \cite{Guerci2019} by direct minimization of the ghost-GA functional \eqn{functional}.\\
The effect of $h\not=0$ on the Mott insulating phase can be anticipated from the zero-field results of the previous section. In that section we found the energy $E(m)$ at fixed magnetization to be smooth and minimum at $m=0$, see Fig.~\ref{fig:Energy_vs_magnetization}. As we switch on $h$, a Legendre transform maps the energy functional to 
\begin{equation}
    E(h) = \min_{m} \left( E(m) - hm \right) \, .
    \label{E(h) vs E(m)}
\end{equation}
For any finite $h$, the minimum of $E(h)$ is displaced to a finite magnetization $\bar{m}(h)$, see the curve $E(h=0.01)$ in Fig.~\ref{fig:Energy_vs_magnetization}. The system is therefore a partially polarized Mott insulator, with a finite magnetic susceptibility at zero field~\cite{Guerci2019, pasqua2026topology, tagliente2025spinons}, 
\begin{equation}
    \chi = \partial_h \bar{m}|_{h\to0} = \big( \partial^2_m E(m)\big)^{-1}_{\bar{m} \to 0}.
\end{equation}
\begin{figure*}[t]
    \centering
    \includegraphics[width=1.0\textwidth]{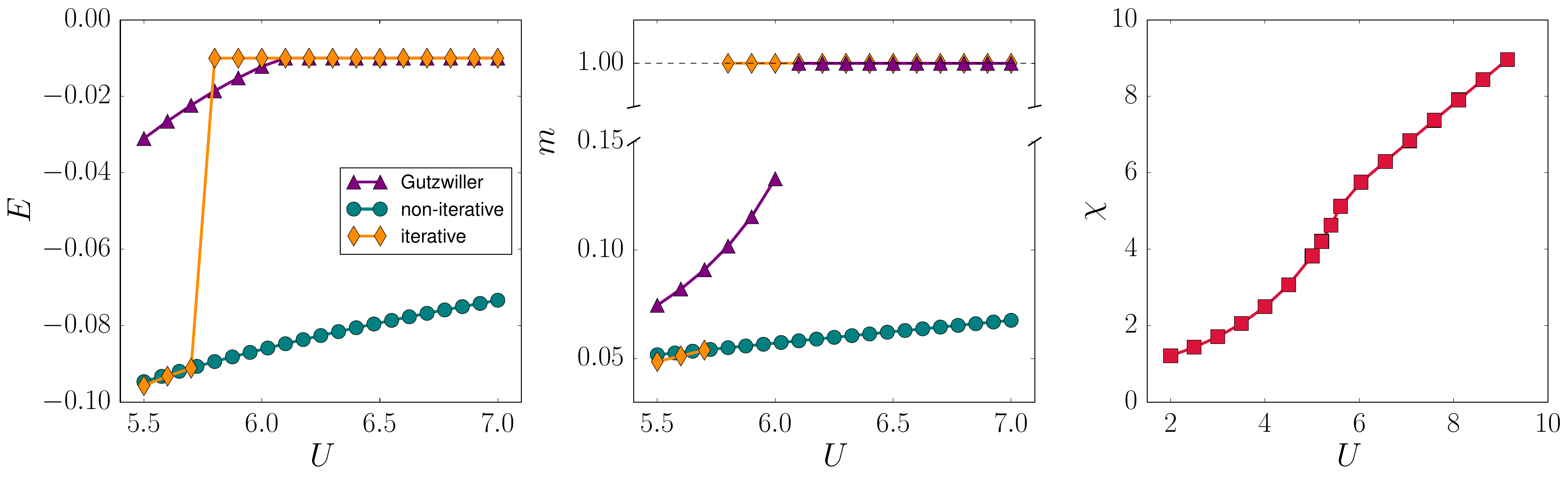}
    \caption{Ghost-GA results for the single-band Hubbard model in a magnetic field $h=0.01$ across the Mott transition, comparing the iterative embedding scheme (\emph{iterative}) with the direct minimization of the ghost-GA functional \eqn{phi Lanczos} (\emph{non-iterative}); the standard Gutzwiller approximation (\emph{Gutzwiller}) is shown for reference. \emph{Left panel}: energy per site $E$ as a function of $U$. The iterative embedding energy displays an unphysical jump and, beyond it, collapses onto the standard GA energy, whereas direct minimization gives a smooth curve with lower energy. \emph{Middle panel}: physical magnetization $m$ as a function of $U$. Both the standard GA and the iterative embedding scheme reach a fully polarized $m=1$ band insulator, while direct minimization remains only partially polarized, $m<1$. \emph{Right panel}: static zero-field spin susceptibility $\chi$ as a function of $U$, obtained from direct minimization in the Mott phase and from the embedding scheme in the metallic phase; $\chi$ stays finite and evolves continuously across the transition.}
    \label{fig:FMresults}
\end{figure*}\\
\noindent
We now examine how the embedding scheme \eqn{Lagrangian} and the direct minimization of the ghost-GA functional \eqn{functional} behave in this regime. We start from the embedding scheme. Naively, we could expect a finite field $h$ to remove the primary difficulty encountered at zero field by lifting the ground state degeneracy. However, the resulting solution is unphysical. As shown in the left panel of Fig.~\ref{fig:FMresults}, the embedding energy has a jump as $U$ increases, which is physically incorrect. Beyond this jump, the solution collapses onto the conventional GA energy, also shown in the figure, and describes a fully polarized, $m=1$, trivial band insulator with energy $E=-h$ per site, see the middle panel of Fig.~\ref{fig:FMresults}. A closely related instability has also been observed in zero-temperature DMFT, where the Mott insulating solution becomes fully polarized in an arbitrarily small magnetic field \cite{Hewson}.\\
The direct minimization of the functional \eqn{functional}, using the wavefunction parametrization \eqn{phi Lanczos}, does not suffer from this instability. The energy curve is smooth, with no discontinuous jump, see the left panel of Fig.~\ref{fig:FMresults}, and the corresponding Mott insulating solution remains only partially polarized with $m<1$, see the middle panel of Fig.~\ref{fig:FMresults}. Its variational energy is obviously lower than the unphysical fully polarized one.\\
\noindent
Finally, the right panel of Fig.~\ref{fig:FMresults} shows the static zero-field spin susceptibility, computed from the direct-minimization solution in the Mott phase and using the embedding scheme in the metallic one. The susceptibility remains finite in the Mott phase and evolves continuously across the transition. This is a considerable improvement over the standard GA, which predicts a diverging susceptibility at the Mott transition, just as DMFT does \cite{Hewson} when one tries to compute the spin susceptibility from energetics instead of using the Bethe-Salpeter equation \cite{DMFT}. We remark that in standard Gutzwiller it is still possible to obtain a finite spin susceptibility in the Mott phase using an RPA approach \cite{michele-fluctuations}, even though in that case the critical $U_c$ is overestimated.
\section{Antiferromagnetic phase}
\begin{figure}[t!]
\centering
\includegraphics[width=0.5\textwidth]{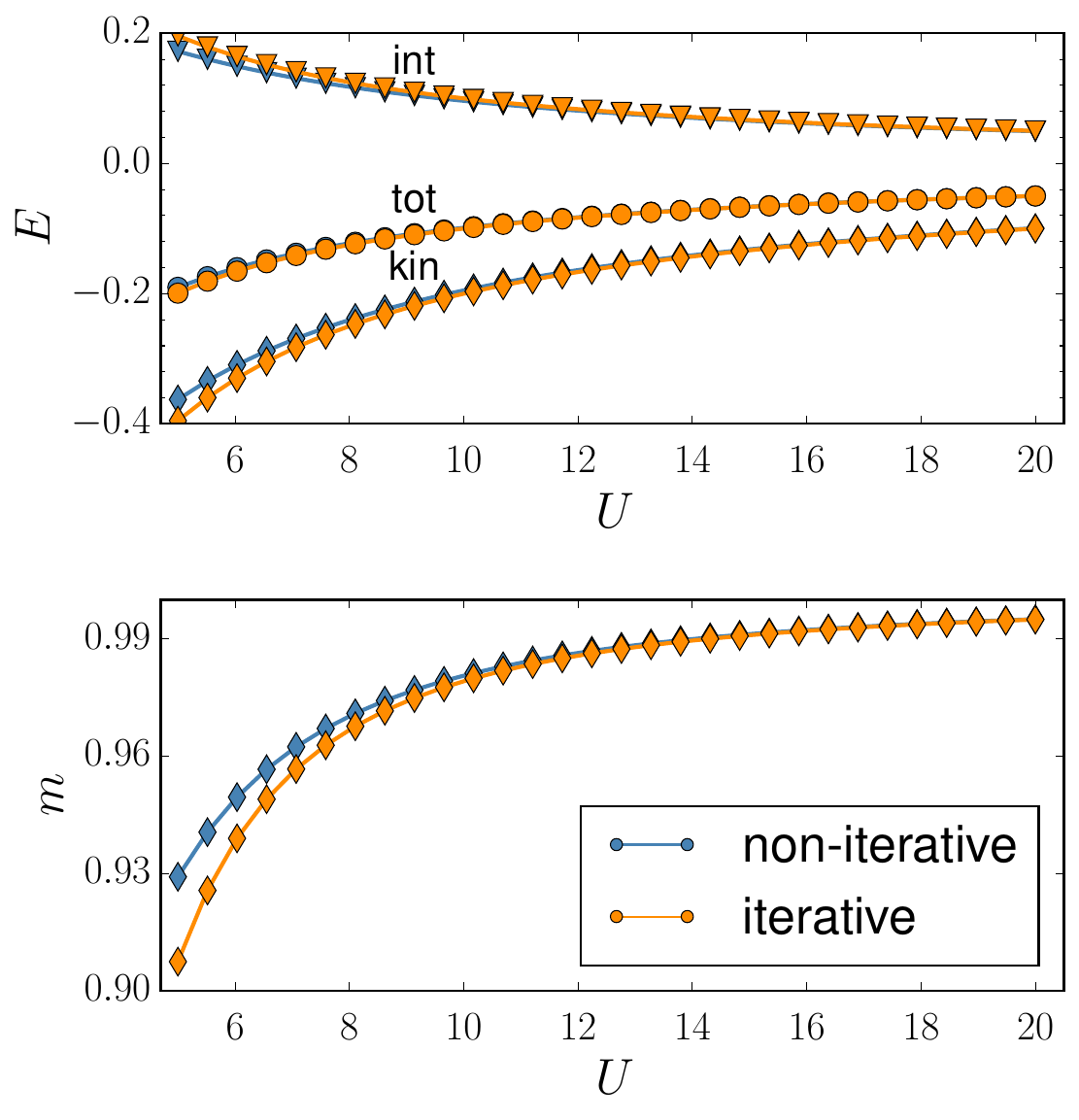}
\caption{Comparison between the energies and the magnetization per site obtained with the iterative embedding scheme (\emph{iterative}) and with the direct minimization of the impurity wavefunction \eqn{phi 1-4 AFM} (\emph{non-iterative}). From top to bottom in the upper panel: local interaction energy (\textit{int}), total energy (\textit{tot}), and kinetic energy (\textit{kin}), as a function of $U$.}
\label{emb_vs_dir_afm}
\end{figure}

\begin{figure*}
\centering
 \centering
    \includegraphics[width = \textwidth]{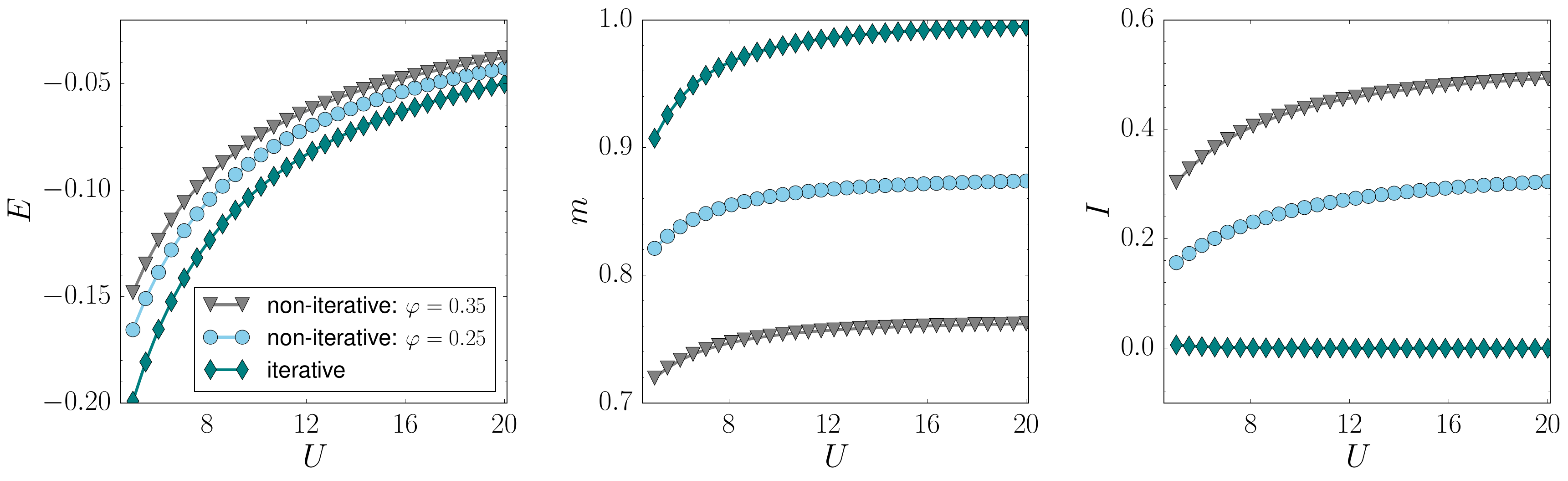}
\caption{Comparison between the true ground state within the \textit{iterative} embedding scheme and the \textit{non-iterative} direct minimization using the impurity wavefunction \eqn{phi 1-4 AFM} at fixed $\varphi$. From left to right: total energy per site, staggered magnetization $m$, and local mutual information $I$ in units of $ \ln 2$ as a function of $U$. Fixing $\varphi$ makes it possible to stabilize more correlated states at the cost of a modest increase in variational energy.}
\label{fixedphivsemb}
\end{figure*}

\noindent
As discussed in the Introduction, the difficulties encountered in iteratively solving the embedding problem can be traced back to the fact that the actual ground state of the half-filled Hubbard model on a bipartite lattice exhibits antiferromagnetic order, which is canted in the presence of a Zeeman field. Consequently, we anticipate that, when we allow for antiferromagnetism, the iterative method should converge effectively. This is precisely what we demonstrate in this section comparing the results of the ghost-GA with DMFT. Furthermore, one can readily adapt the aforementioned minimization scheme in the antiferromagnetic case by considering two distinct impurity wavefunctions for the two sublattices with opposite impurity magnetizations, $m$ and $-m$, and determine the minimum with respect to $m$. In this case, we relax the constraint that the Hubbard bands are unpolarized so that 
$\ket{\phi_n}$, $n=1,\dots,4$, in \eqn{phi 1-4} become now
\beal
\ket{\phi_1} &= \fract{\sin\varphi}{\sqrt{2}}\,\Big(
\ket{2,\down,0}\ket{\up}-\ket{0,2,\down}\ket{\up}\Big)\\
&\qquad -\fract{\cos\varphi}{\sqrt{2}}\,\Big(\ket{2,\up,0}\ket{\down}
-\ket{0,2,\up}\ket{\down}\Big)\,,\\
\ket{\phi_2} &= \fract{\sin\varphi}{\sqrt{6}}\,\Big(\ket{\up,\down,\up}\ket{\down}+\ket{\up,\up,\down}\ket{\down}\\
&\qquad\qquad \qquad\qquad\qquad
-2\ket{\up,\down,\down}\ket{\up}\Big) \\
& \qquad + \fract{\cos\varphi}{\sqrt{6}}\,\Big(\ket{\down,\down,\up}\ket{\up}
+ \ket{\down,\up,\down}\ket{\up}\\
&\qquad\qquad \qquad\qquad\qquad\qquad
-2 \ket{\down,\up,\up}\ket{\down}\Big)\,,\\
\ket{\phi_3} &= \fract{\sin\varphi}{\sqrt{2}}\,\Big(\ket{2,0,\up}\ket{\down}-\ket{0,\up,2}\ket{\down}\Big)\\
&\qquad 
+\fract{\cos\varphi}{\sqrt{2}}\,\Big(\ket{0,\down,2}\ket{\up}-\ket{2,0,\down}\ket{\up}\Big)\,,\\
\ket{\phi_4} &= -\cos\varphi \ket{\up,0,2}\ket{\down} + \sin\varphi \ket{\down,0,2}\ket{\up}\,,
\label{phi 1-4 AFM}
\eal
while $\ket{\phi_0}$ equals \eqn{phi-0}. The two wavefunctions corresponding to opposite magnetization 
are identified by $\varphi\in [0,\pi/4]$ and $\pi/2-\varphi$. 
The direct optimization using such impurity wavefunction, although now less accurate than the iterative embedding scheme, as we show, possesses the advantageous feature of describing states forced to have fixed and arbitrary order parameter, the staggered magnetization $m$. This is not feasible in the iterative schemes that directly provide the ground state value of $m$.
\begin{figure}
\centering
    \centering
    \includegraphics[width=0.5\textwidth]{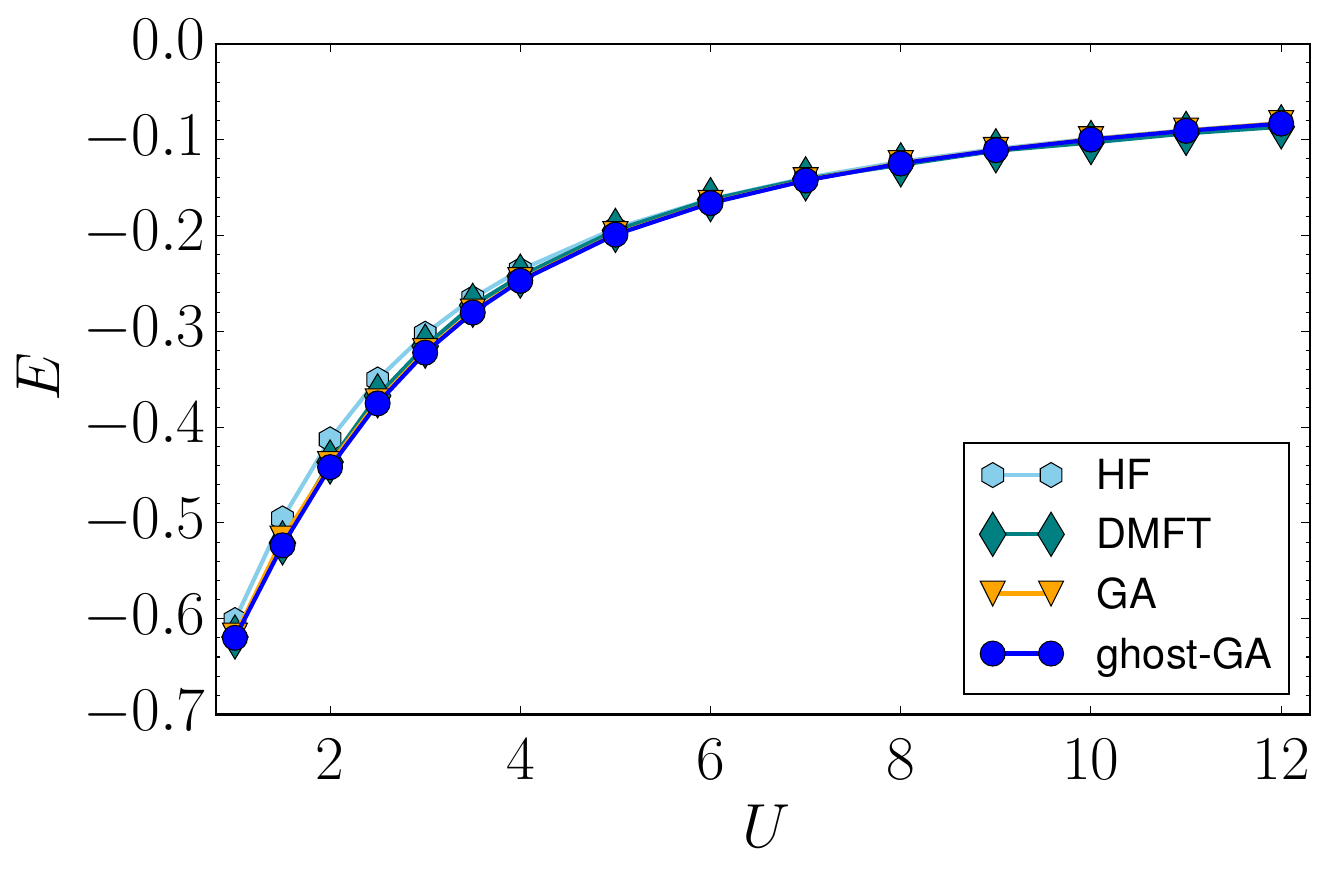}
    \caption{Total energy $E$ versus $U$ for the antiferromagnetic phase in the single-band Hubbard model, calculated using HF, DMFT, GA, and ghost-GA. At small $U$, the ghost-GA and DMFT energies are noticeably lower than those of HF and GA, whereas the differences become negligible at large $U$. DMFT data courtesy of Adriano Amaricci.    
    }
    \label{dmftvsgga} 
\end{figure}\\
\noindent
In Fig.~\ref{dmftvsgga} we compare the iterative ghost-GA results with the DMFT 
ones \cite{Adriano}, which are obtained using exact diagonalization as impurity solver \cite{edipack1, edipack2}.
The agreement is excellent, and both energies do not differ much from conventional GA as well as Hartree-Fock (HF), as noted in \cite{Giorgio, gabriele}. In other words, the ground state properties in high dimensions are already well captured by HF, while dynamical ones may differ substantially \cite{Giorgio,Taranto}.\\ Having ascertained that the embedding scheme offers the same physical picture as DMFT, we proceed to study the antiferromagnetic phase through the variational minimization using the impurity wavefunction \eqn{phi Lanczos} with \eqn{phi-0} and \eqn{phi 1-4 AFM}. The results of this method with those obtained by the embedding scheme are compared in  Fig.~\ref{emb_vs_dir_afm}. We observe that the energy and magnetization of the embedding solution are lower than those obtained by the direct minimization, and become equal only at strong coupling.
\\
\noindent
The evidence that the DMFT, GA and ghost-GA values of $m$ approach the maximum $m=1$ value for large $U$, akin to HF, is a consequence of the inadequate description of spatial fluctuations of these methods that rely on the limit of infinite lattice coordination. In reality, the staggered magnetization saturates to a smaller value as the lattice dimensionality decreases, thereby amplifying the effects of quantum fluctuations. A simple way to quantify the role of quantum fluctuations   
is the mutual information 
\begin{equation}
    I = S(\rho_\up) + S(\rho_\down) - S(\rho) \, ,
\end{equation}
where $\rho$ is the local reduced density matrix, $\rho_\sigma$ is the spin-resolved one, and $S(\rho)$ the von Neumann entropy. This quantity is equivalent
to the nonfreeness \cite{gabriele} and measures the distance of the state from a Gaussian one. 
It has been shown \cite{gabriele} that local embedding schemes, when allowed to break symmetry, 
generally reduce to a Hartree-Fock description in the strong-coupling limit with the mutual information $I$ close to zero. The minimization of the ghost-GA functional \eqn{functional} employing the impurity wavefunction \eqn{phi Lanczos} with components \eqn{phi-0} and \eqn{phi 1-4 AFM} presents a unique opportunity to access variationally states characterized by arbitrary staggered magnetization $m$, which is determined by the parameter $\varphi$. In Fig.~\ref{fixedphivsemb}, we present the total energy per site, magnetization, and mutual information $I$ of states optimized at fixed $\varphi$ in comparison with the solution obtained within the embedding scheme. Notably, despite a relatively modest energy difference, the staggered magnetization $m$ and, particularly, the mutual information, can exhibit substantial variations.   

\section{Conclusion}
\noindent
The ghost-Gutzwiller variational approach within the Gutzwiller approximation can be formulated in a quantum embedding scheme that comprises two self-consistently coupled variational wavefunctions. The first is the wavefunction of an impurity model, while the second is a Slater determinant ground state of a non-interacting periodic Hamiltonian with parameters that are non-linear functional of the impurity wavefunction. The self-consistency condition establishes a connection between the local one-body density matrix of the Slater determinant and the reduced one-body density matrix of the bath in the impurity wavefunction. The most straightforward and efficient approach to solving the variational problem is by introducing a set of Lagrange multipliers that correspond to specific components of the reduced one-body density matrices. This allows the problem to be recast into two separate and self-consistently coupled eigenvalue problems, which can be readily solved iteratively.\\
\noindent
We have demonstrated that this seemingly legitimate reformulation of the ghost-Gutzwiller variational problem may occasionally yield incorrect results. The flaw primarily arises when the physical Hamiltonian contains terms that split the degeneracy of the atomic limit pertinent to the Mott insulator. In the conventional Gutzwiller approximation, this circumstance leads to a first-order metal-insulator transition into a Mott phase where each site is frozen in the lowest-energy atomic configuration. In the ghost-Gutzwiller approximation, the Mott insulator remains unchanged, but the metal-insulator transition is accompanied by a finite jump in the variational energy, which is unphysical. We have shown that the correct results can be achieved by treating the complete impurity wavefunction as a variational object without requiring it to be the ground state of an Anderson impurity model. While this approach exhibits reduced efficiency compared to the iterative method, it presents significant advantages. Notably, it allows a straightforward enforcement of symmetries that may be broken in the actual ground state, thereby enabling the exploration of metastable states with intriguing properties, such as spin-charge fractionalization~\cite{tagliente2025spinons,pasqua2026topology,pasqua2026FLstar} or spin-liquid insulators. On the contrary, when we allow for symmetry breaking and the insulator is describable in mean-field, the iterative scheme is very efficient and provides sensible results. 

\section{Acknowledgments}
\noindent
We are deeply grateful to Adriano Amaricci for providing us with the DMFT data for the antiferromagnetic phase. We also acknowledge insightful discussions with G. Kotliar, N. Lanat\`a, C. Mejuto Zaera, and G. Sangiovanni


\begin{thebibliography}{35}
\bibitem{Adriano} The {DMFT} data have been kindly provided to us by {Adriano} {Amaricci}.
\end{thebibliography}
\end{document}